\documentclass{czjphys}         
\usepackage{graphicx}
\begin{document}

\title{Photon air showers at ultra-high energy
and the photonuclear cross-section
\footnote{Based on a talk presented
at the international conference ``From Colliders to Cosmic Rays'',
Prague, September 7-13 (2005)}
}

%


\authori{
M.~Risse$^{a,b}$\footnote{Electronic address: markus.risse@ik.fzk.de},
P.~Homola$^b$, R.~Engel$^a$, D.~G\'ora$^b$, D.~Heck$^a$,
J.~P\c{e}kala$^b$, B.~Wilczy\'nska$^b$, H.~Wilczy\'nski$^b$}
\addressi{$^a$ Forschungszentrum Karlsruhe, Institut f\"ur Kernphysik,
76021~Karlsruhe, Germany
\\
$^b$ Institute of Nuclear Physics, Polish Academy of Sciences,
ul.~Radzikowskiego 152, 31-342 Krak\'ow, Poland}

\authorii{} \addressii{}
%
\headauthor{M.~Risse et al.}   
\headtitle{Photon air showers and photonuclear cross-section}
\lastevenhead{M.~Risse et al.: Photon air showers at highest energies
and the photonuclear cross-section}
\refnum{A}
\daterec{XXX}    
\issuenumber{0}  \year{2001}
\setcounter{page}{1}
\maketitle
\begin{abstract}
Experimental conclusions from air shower observations on cosmic-ray
photons above $10^{19}$~eV
are based on the comparison to detailed shower simulations. 
For the calculations, the photonuclear
cross-section needs to be extrapolated over several orders of magnitude
in energy. The uncertainty from the cross-section extrapolation 
translates into an uncertainty of the predicted shower features for
primary photons and, thus, into uncertainties for a possible data 
interpretation. After briefly reviewing the current status of
ultra-high energy photon studies,
the impact of the uncertainty of the photonuclear
cross-section for shower calculations is investigated.
Estimates for the uncertainties in the main shower
observables are provided.
Photon discrimination is shown to be possible even for rapidly
rising cross-sections.
When photon-initiated showers are identified,
it is argued that the sensitivity of photon shower 
observables to the photonuclear cross-section can in turn be
exploited to constrain the cross-section at energies not accessible
at colliders.
\end{abstract}

\section{Introduction}  
\label{sec-intro}

Photons around and above $10^{19}$~eV might provide a key to
understanding the origin of cosmic rays.
Substantial fluxes of these ultra-high energy (UHE) photons
are predicted in top-down models of cosmic-ray origin.
In addition, UHE photons are produced by the
GZK process of resonant photoproduction of pions~\cite{gzk},
in analogy to GZK neutrinos.

Experimentally, UHE photons can be discerned from nuclear
primaries due to differences in the expected shower signatures.
So far, no claim of a photon detection exists and upper limits
to UHE photons were set. Any conclusions about UHE photons
rely on the comparison of data to detailed simulations of 
photon-induced showers.
Although these photon showers are dominated by electromagnetic
interactions, a source of uncertainty is the photonuclear cross-section,
which has to be extrapolated over several orders of magnitude
in energy from laboratory data for calculating showers induced by 
UHE photons.

In this work, after giving an overview of the current status of
UHE photon studies, both the impact of the uncertainty from
the photonuclear cross-section to photon shower features
and possible prospects for constraining the cross-section are
investigated.
The plan of the paper is as follows.
In Section~\ref{sec-cr}, scenarios for producing cosmic-ray
photons above $10^{19}$~eV = 10~EeV are briefly described.
Features of showers induced by photons are discussed in 
Section~\ref{sec-eas}.
In Section~\ref{sec-data}, the experimental situation concerning upper
limits to photons is summarized. 
The role of the photonuclear cross-section is investigated in
Section~\ref{sec-sigma}.

\section{Cosmic-ray photons}
\label{sec-cr}

Motivated in particular by reports from the AGASA experiment
about a possible continuation of the cosmic-ray energy spectrum
without a GZK cutoff,
many top-down scenarios were proposed~\cite{bhat-sigl}.
In these models, cosmic rays are produced by the decay or
annihilation of superheavy dark matter~\cite{shdm} or
topological defects~\cite{td}.
A common feature of these models is the prediction of a
significant flux of photons arriving at Earth.
Also in the Z-burst model~\cite{zb}, large photon fractions are
predicted.
Stringent upper limits to photons would disfavour many realizations
of these top-down models, although some uncertainty exists in
calculating the propagation of UHE photons~\cite{sarkar03}.

%
\bfg[t]                     
\bc                         
\includegraphics[width=0.50\textwidth]{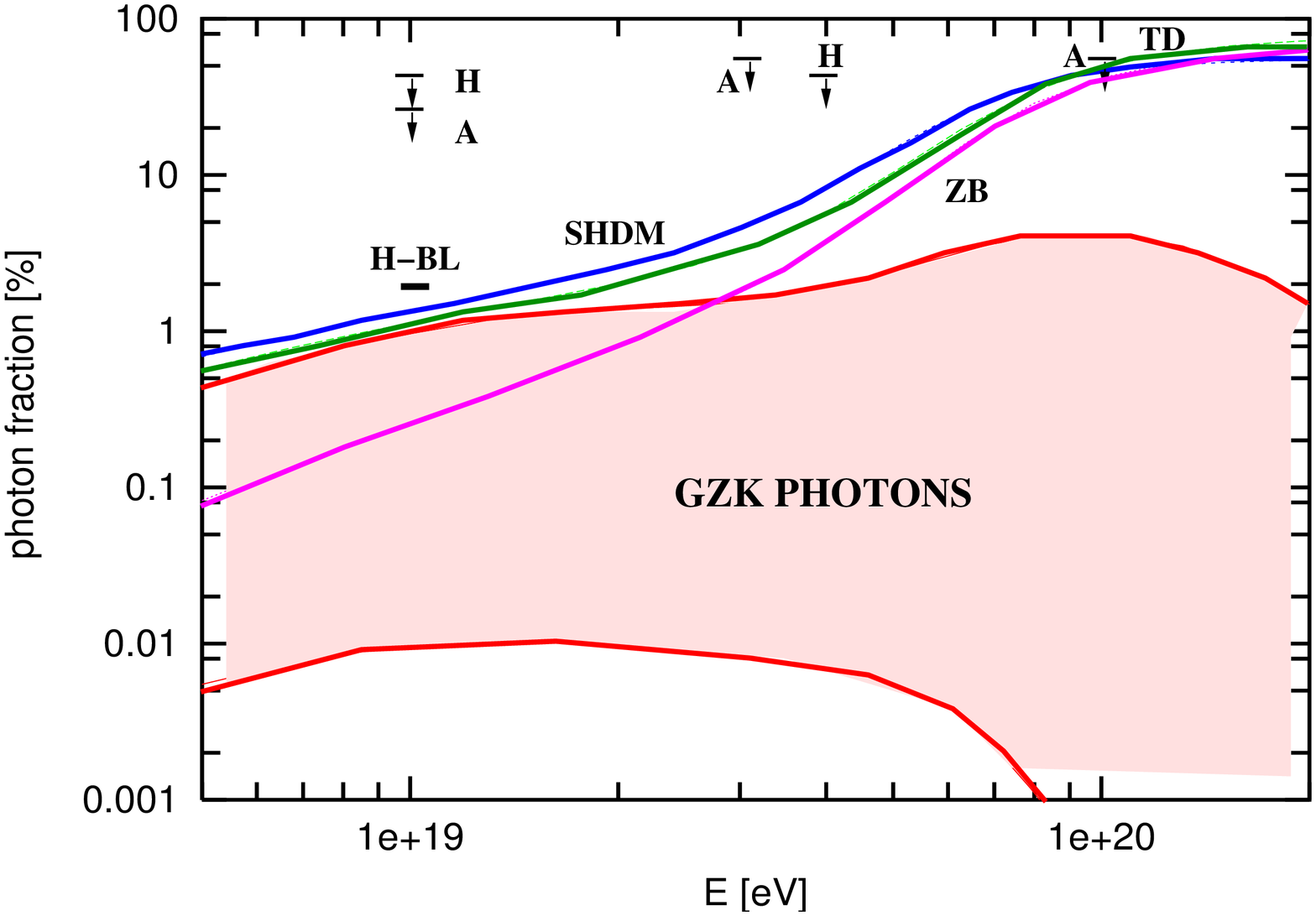}
\includegraphics[width=0.48\textwidth]{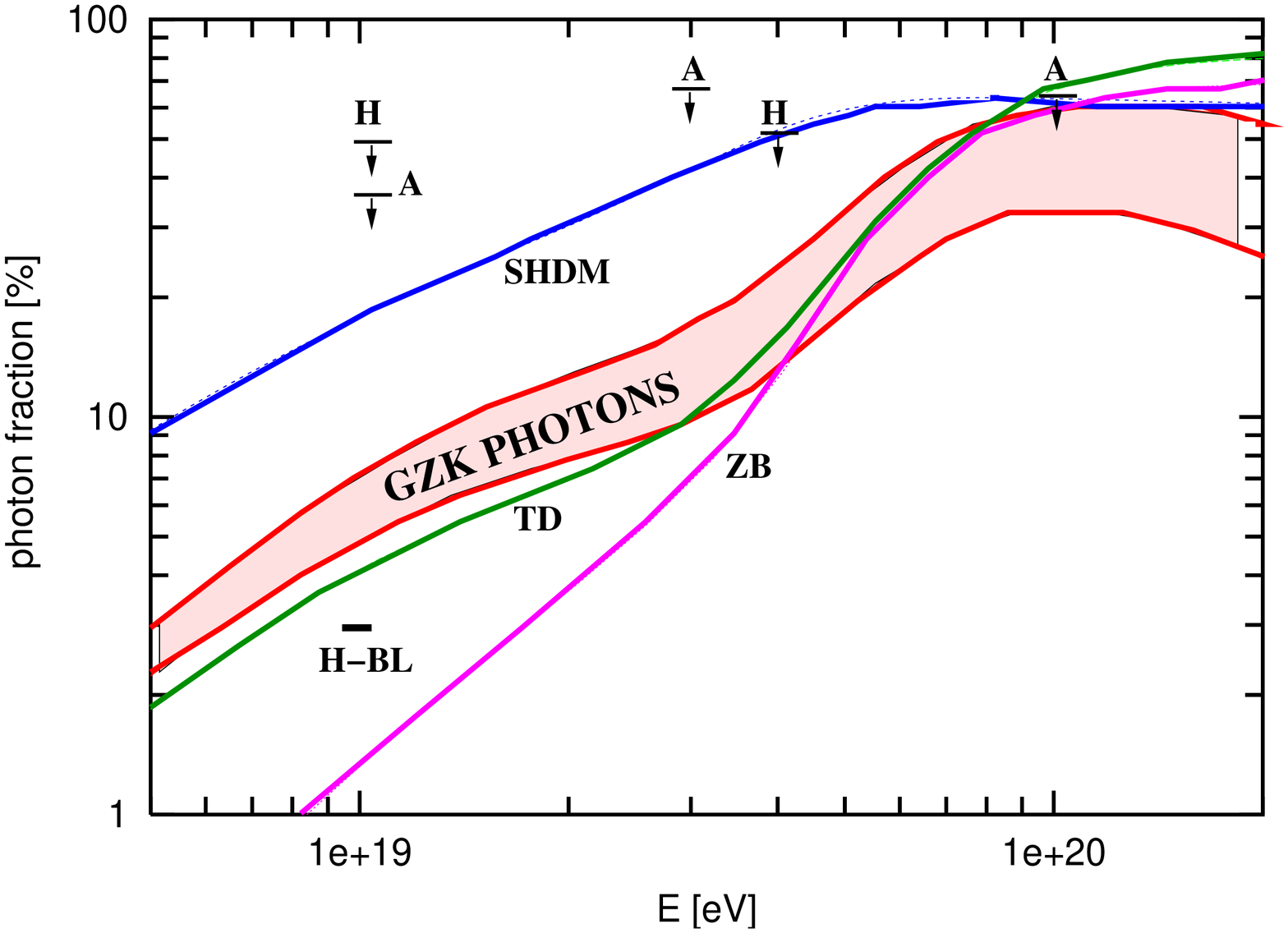}

\ec                         
\vspace{-2mm}
\caption{Production of photons by the GZK process: Bottom-up sources
were fitted to the HiRes spectrum (left) and the AGASA spectrum 
(right). Plotted is the resulting photon fraction, integrated above the
primary energy, as a function of the primary energy threshold.
The ranges shown
for the GZK photons come from varying the source and background
parameters. Figures are taken from~\cite{models}.}
\label{fig-gzk}
\efg                        

Even in ``conventional'' bottom-up models, in which nuclear particles
are accelerated in astrophysical sites to highest energies,
UHE photons are expected. They arise from the GZK process during
propagation.
Such GZK photons were studied in detail recently in~\cite{models}.
In Figure~\ref{fig-gzk}, results of these calculations are shown:
For various assumptions on source parameters (distribution, emission
features) and on radio backgrounds and magnetic fields important for
photon propagation, fits were performed both to the
HiRes (cutoff-like)~\cite{hires-gzk}
and AGASA (no-cutoff-like)~\cite{agasa-gzk} cosmic-ray spectra.

In case of the HiRes spectrum, the range of possible photon fractions
extends from almost negligible ($<$0.01\%) to values of 1\% above
10~EeV and 5\% above 100 EeV. For the completed Pierre Auger Observatory
(Northern and Southern site)~\cite{auger}, the latter values would imply
event rates of up to $\simeq$60 photons/year above 10~EeV and
$\simeq$2-3 photons/year above 100 EeV.
Such a signal seems detectable; in turn, an absence of these
GZK photons, and correspondingly an upper limit to photons,
might translate into constraints for bottom-up scenarios.

In case of the AGASA spectrum, it turned out that even bottom-up 
scenarios might allow a fit to a no-cutoff spectrum.
The GZK photon fraction is then very large, however.
(The term ``GZK photons'' refers to the way these photons are produced;
it does not imply that the observed energy spectrum necessarily
exhibits a GZK cutoff.)
Thus, a connection between the shape of the energy spectrum and the
photon fraction predicted in bottom-up models exists~\cite{models}.
It seems worthwhile to note that in turn, a small-enough photon limit
(e.g.~below 2-3\% above 10~EeV or below 10-20\% above 100~EeV)
would then disfavour a no-cutoff-like spectrum for the scenarios
regarded in~\cite{models}, unless specific new physics such as
Lorentz invariance violation is invoked (of course, top-down models
would be constrained by the same photon limit).

\section{Air showers induced by photons}
\label{sec-eas}

%
\bfg[t]                     
\bc                         
\includegraphics[width=0.8\textwidth]{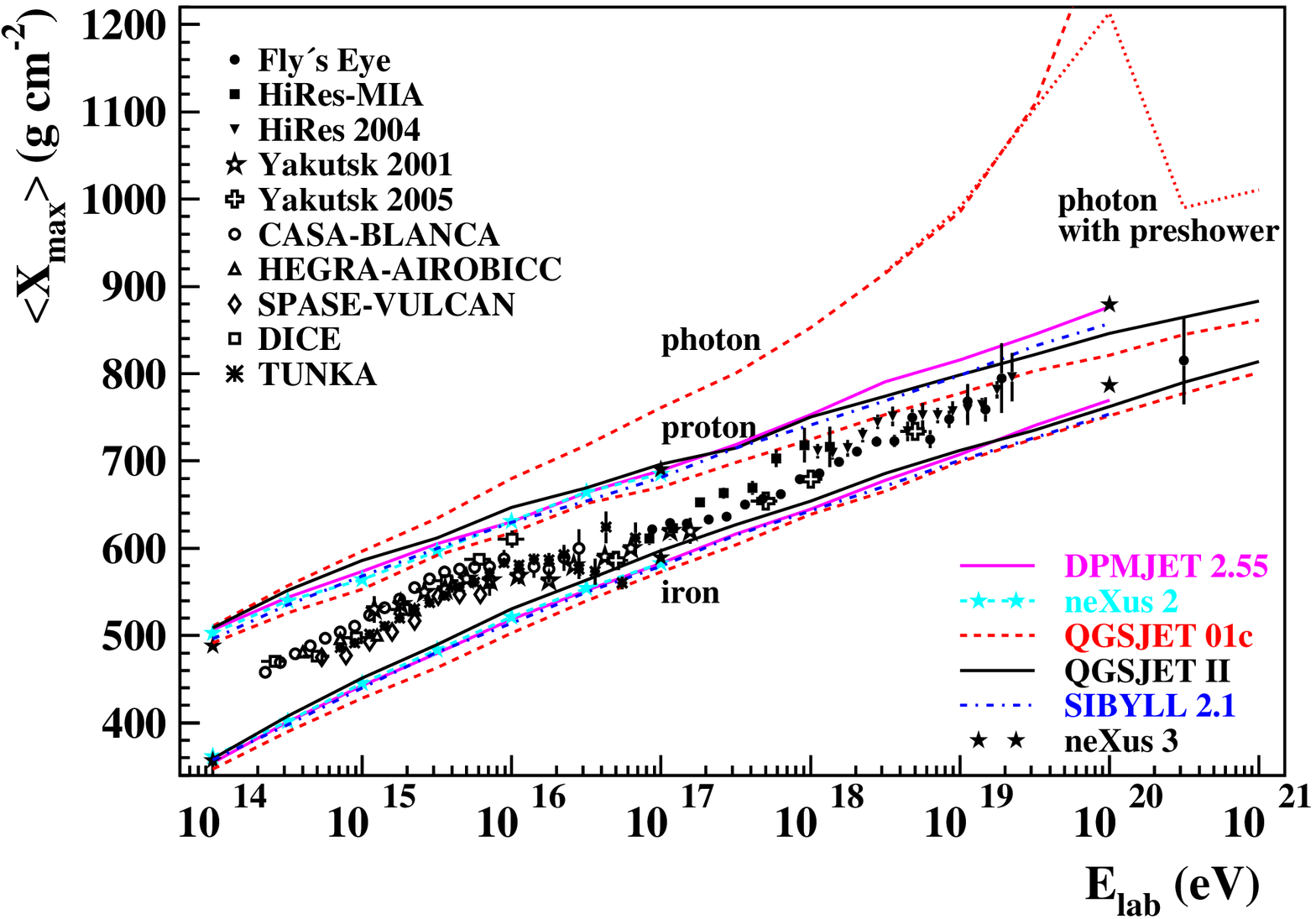}
\ec                         
\vspace{-2mm}
\caption{Average depth of shower maximum $X_{\rm max}$ versus primary
energy simulated
for primary photons, protons and iron nuclei. For nuclear primaries,
calculations for different hadronic interaction models are shown.
Also shown are experimental data. For references to the experiments and
interaction models see~\cite{xmax-heck}. }
\label{fig-xmax}
\efg                        

In Figure~\ref{fig-xmax}, the average depth of shower maximum 
$X_{\rm max}$ is shown
vs.~primary energy as calculated for different primary particles. The
shower maxima of primary photons are separated from those of nuclear
primaries by $\simeq$200~g~cm$^{-2}$ or more at 10 EeV.
The elongation rate (slope of the curve) is larger for photons
even at relatively small energies.
This elongation rate for photons can be understood within the
Heitler toy model of shower development~\cite{heitler}.
The slope is increased above a few EeV with the 
LPM effect~\cite{lpm} becoming increasingly important.
At even higher energy (above 50~EeV, depending on geomagnetic field),
UHE photons may convert in the geomagnetic field and
form a preshower~\cite{erber} before entering the atmosphere.

%
\bfg[t]                     
\bc                         
\includegraphics[width=0.45\textwidth]{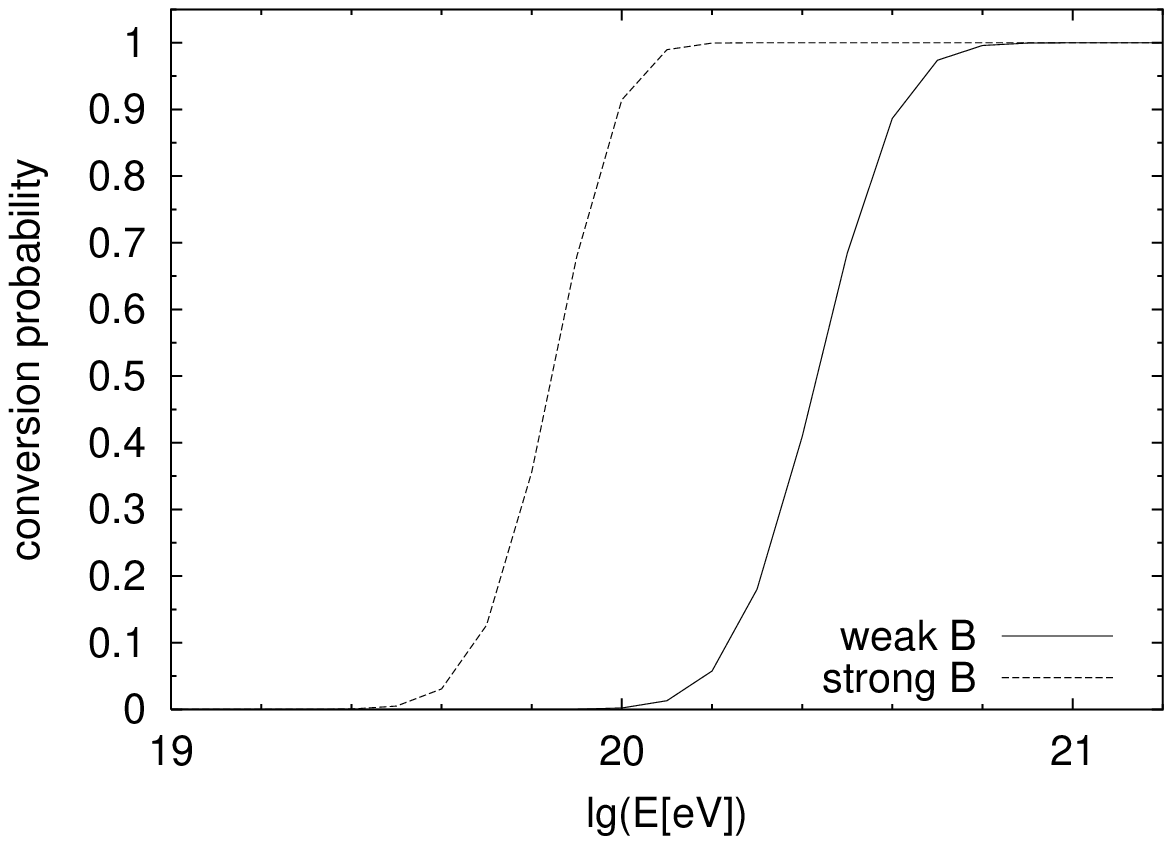}
\includegraphics[width=0.52\textwidth]{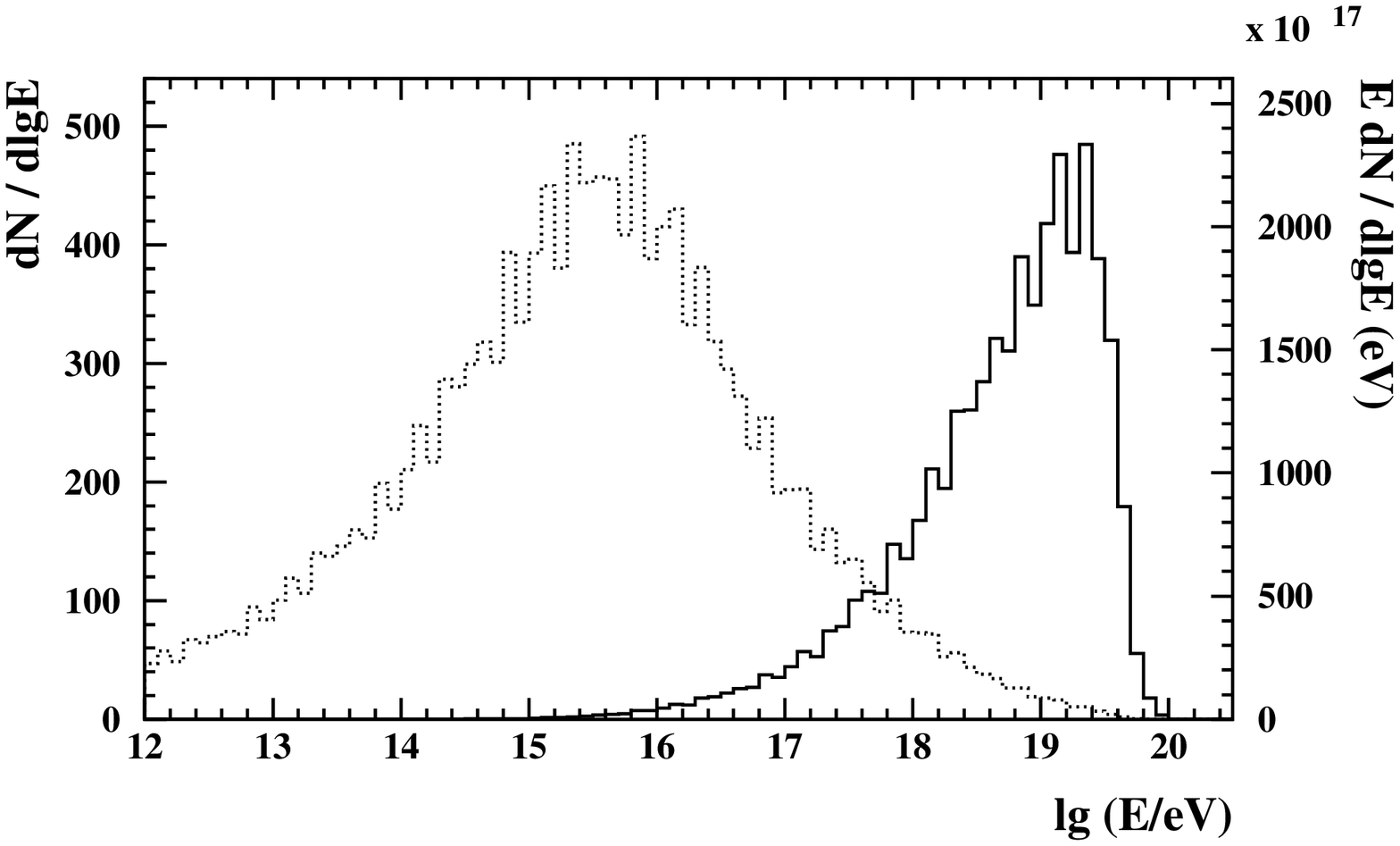}
\ec                         
\vspace{-2mm}
\caption{Preshower features: (Left) Conversion probability versus
primary energy for two different arrival directions: Small angle
(``weak B'') and almost perpendicular (``strong B'') to the local
magnetic field.
Calculation performed for the Auger Southern site~\cite{homola}.
(Right) Spectrum of preshower particles (with and without energy 
weighting) on top of the atmosphere calculated for photons for the
geometry of the 320~EeV Fly's Eye event~\cite{fe04}. }
\label{fig-conv}
\efg                        

The directional dependence of the preshower effect is illustrated
in Figure~\ref{fig-conv} (left). In this example, photons of
100~EeV primary energy do not convert when entering almost parallel
to the local geomagnetic field lines. 
For a more perpendicular incidence, in most cases a preshower is
formed~\cite{homola}.

If a preshower is formed, the energy is distributed among secondary
electrons, positrons and, mostly, photons.
As a specific example, in Figure~\ref{fig-conv} (right) the energy
spectrum of preshower particles on top of the atmosphere is shown.
As the original particle, a primary photon was assumed for the
geometry of the 320~EeV Fly's Eye event~\cite{bird}.
On average, about 1400 lower-energy
particles are produced. Most energy is carried by particles
of $\simeq$10~EeV energy, i.e.~about 1.5 orders of magnitude below
the original primary energy~\cite{fe04}.

The impact of the LPM and preshower effect is shown for the example of
the highest energy Fly's Eye event in Figure~\ref{fig-profiles}.
Assuming primary photons, the longitudinal profiles are shown for
calculations with both
effects neglected, with only LPM switched on, and for the complete
treatment. Due to the extreme energy of the Fly's Eye event, differences
between the resulting profiles are enormous. While the LPM effect
delays the development and increases fluctuations, the preshower
effect counteracts.

%
\bfg[t]                     
\bc                         
\includegraphics[width=0.48\textwidth]{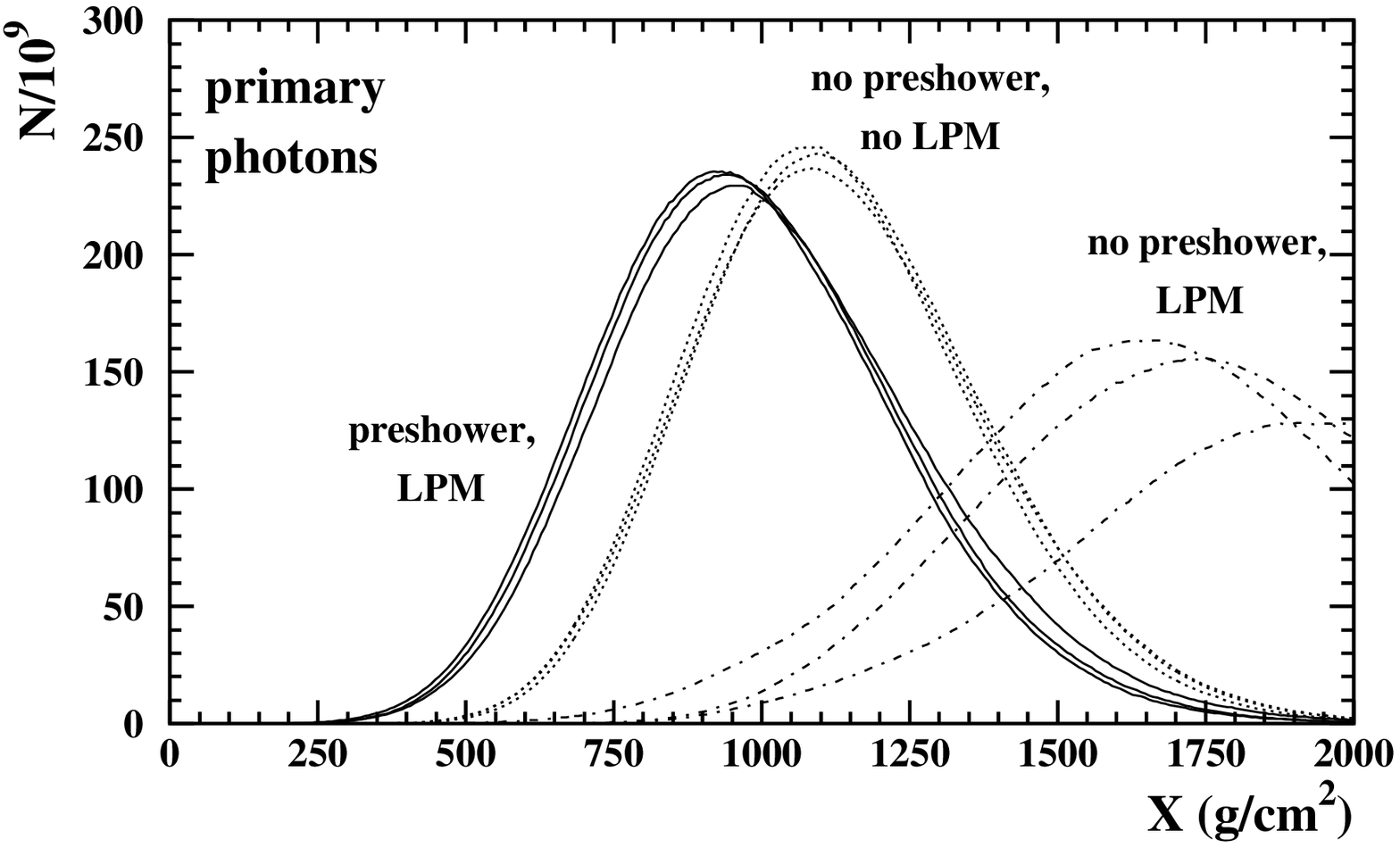}
\includegraphics[width=0.48\textwidth]{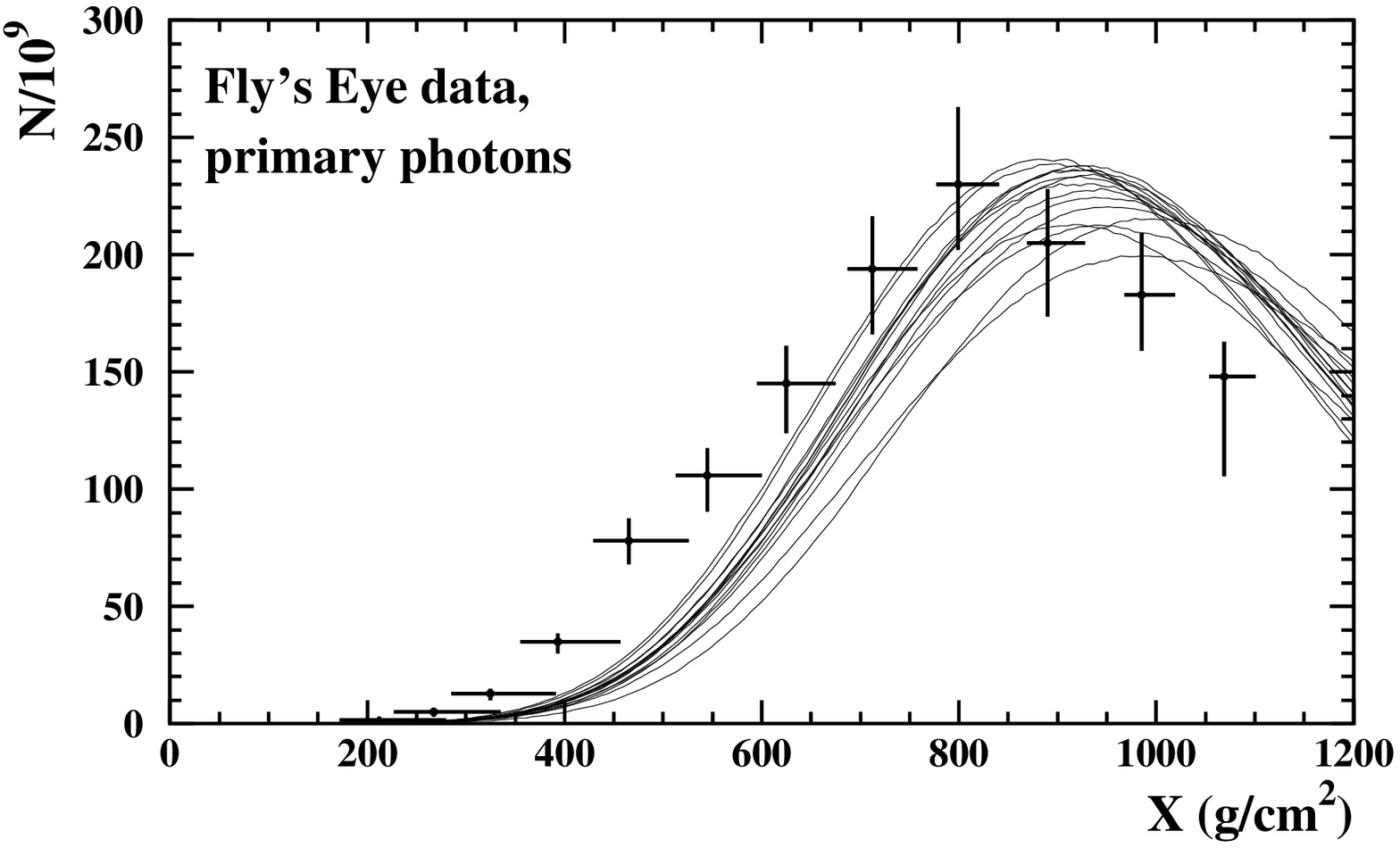}
\ec                         
\vspace{-2mm}
\caption{(Left) Photon shower profiles for the geometry of the 320~EeV
Fly's Eye event for the full simulation (solid), when switching off
the preshower effect (dotted-dashed), and when switching off also
the LPM effect (dotted)~\cite{fe04}.
(Right) Observed profiles of the 320~EeV Fly's Eye event~\cite{bird}
compared to the full simulation for primary photons~\cite{fe04}.
Data points are correlated with respect to shifts in atmospheric depth
$X$.}
\label{fig-profiles}
\efg                        

\section{Data}
\label{sec-data}

For the 320~EeV Fly's Eye event, the observed profile~\cite{bird}
differs from the primary photon prediction by about 1.5 standard
deviations~\cite{halzen_fe,fe04}. 
The comparison is shown in Figure~\ref{fig-profiles} (right).
A photon origin of the Fly's Eye event can not be excluded; profiles
calculated for nuclear primaries fit the data better,
however~\cite{fe04}.

Limits to photons were set by different experiments.
Comparing rates of nearly vertical showers to inclined ones, upper
limits to the photon fraction of 48\% above 10~EeV and 50\% above
40~EeV (95\% CL) were deduced from Haverah Park data~\cite{havpark}.
Based on an analysis of muons in air showers observed by the
Akeno Giant Air Shower Array (AGASA), upper limits were estimated to
be 28\% above 10~EeV and 67\% above 32~EeV (95\% CL)~\cite{agasa}.
In a dedicated analysis of the highest-energy AGASA events, an upper
limit of 67\% above 125~EeV (95\% CL) was set~\cite{risse-prl}.
These limits came from ground arrays. Recently, using $X_{\rm max}$
observed by fluorescence telescopes in hybrid events, a limit of
26\% above 10~EeV (95\% CL) was obtained by the Pierre Auger 
Collaboration~\cite{augericrc}.
These upper limits are compared to some model predictions in
Figure~\ref{fig-uplim}.

%
\bfg[t]                     
\bc                         
\includegraphics[width=0.8\textwidth]{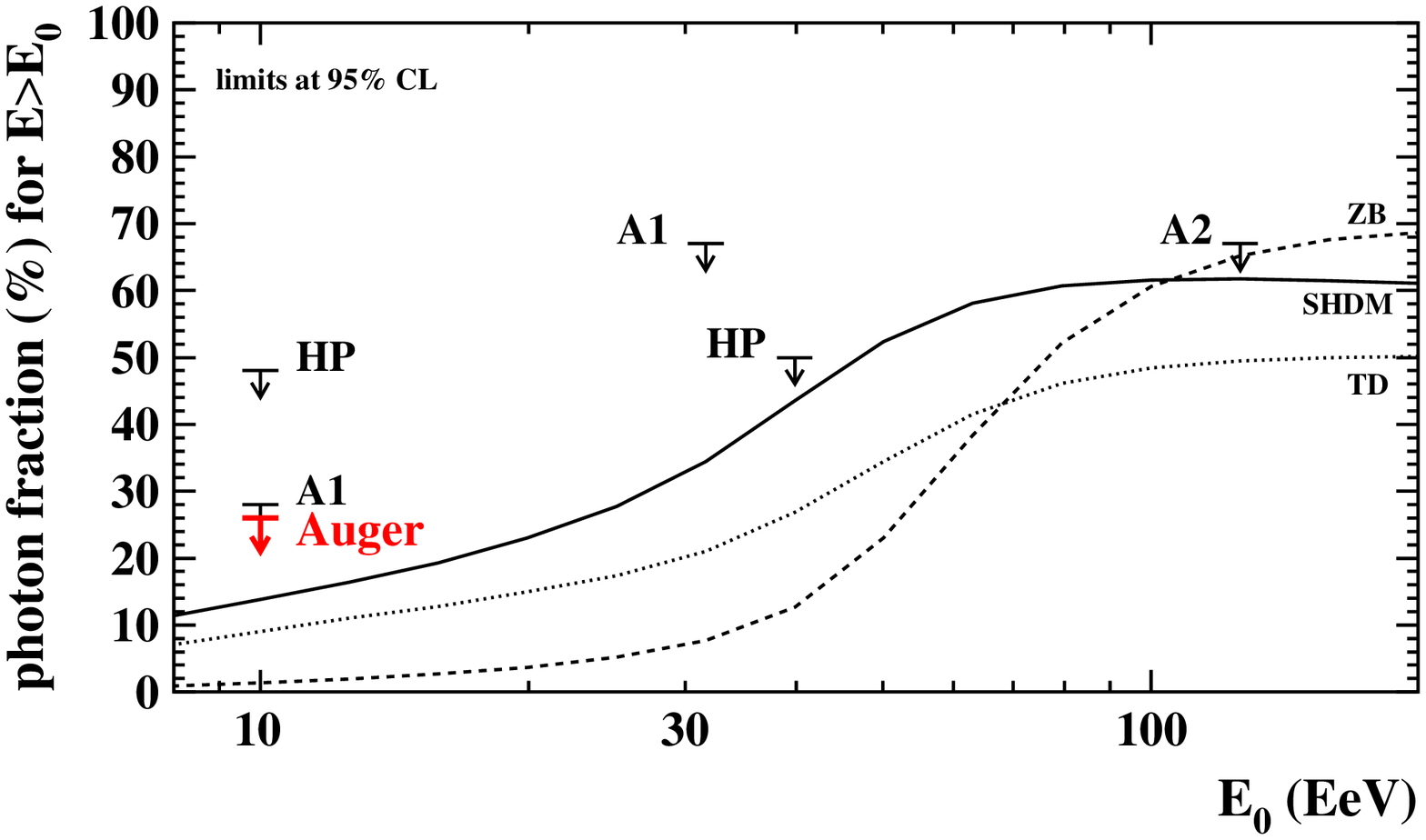}
\ec                         
\vspace{-2mm}
\caption{
Upper limits (95\% CL) to cosmic-ray photon fraction
derived from Haverah Park (HP)~\cite{havpark} and AGASA 
(A1)~\cite{agasa}, (A2)~\cite{risse-prl} data and by
the Auger Observatory~\cite{augericrc}, compared to some
estimates based on top-down models~\cite{models}
(Figure taken from~\cite{augericrc}).}
\label{fig-uplim}
\efg                        

With increased event statistics, much stronger constraints on the
photon flux or the discovery of UHE photons can be expected.
There exists a minimum possible value for an upper limit
that ideally could be reached for a given event statistics.
This is due to the fact that,
assuming a fraction $F_\gamma$ of photons in the primary flux,
a set of $n_{\rm m}$ primaries picked at random is
expected to {\it ab initio} contain no primary photon with
probability $(1-F_\gamma)^{n_{\rm m}}$.
The relation between the minimum possible fraction
$F_\gamma^{\rm min}$ that could be excluded for a given event number
$n_{\rm m}$ (or in turn:
the minimum event number $n_{\rm m}^{\rm min}$ required to
possibly exclude a certain fraction $F_\gamma$) is given by
\begin{equation}
F_\gamma^{\rm min}  = 1-(1-\alpha)^{1/n_{\rm m}}~,~{\rm and}~~~~
n_{\rm m}^{\rm min} = \frac{\ln(1-\alpha)}{\ln(1-F_\gamma)}~,
\end{equation}
with $\alpha$ being the confidence level of rejection.
This theoretical limit is reached only if for each individual shower,
the observations allowed us to completely rule out photons as
primary particle.
Some numerical examples are listed in Tab.~\ref{table}.
\begin{table}[t]
\begin{center}
\caption{Numerical examples for the relation 
$n_{\rm m} \leftrightarrow F_\gamma^{\rm min}$
for the minimum fraction
$F_\gamma^{\rm min}$ that could be
excluded with $n_{\rm m}$ events (or: the minimum number of events
$n_{\rm m}^{\rm min} \leftrightarrow F_\gamma$ 
required to exclude a fraction $F_\gamma$)
for a confidence level $\alpha$ = 95\%.
}
\label{table}
\begin{tabular}{c}
\\
\hline
6$\leftrightarrow$39.3\% ~
10$\leftrightarrow$25.9\% ~
30$\leftrightarrow$9.5\% ~
100$\leftrightarrow$3.0\% ~
300$\leftrightarrow$1.0\% ~
1000$\leftrightarrow$0.3\%
\\
\hline
\end{tabular}
\end{center}
\end{table}

\section{Photonuclear cross-section}
\label{sec-sigma}

The experimental results for UHE photons are based on comparisons
to photon shower simulations. For the simulation, an extrapolation
of the photonuclear cross-section is needed.
This is in particular important for the production of secondary muons
in primary photon showers.
For hadron primaries, most muons are produced in the decay of pions
and kaons generated in collisions of shower hadrons with the target
air nuclei.
In photon-induced showers, however, photoproduction is the main process
that transfers energy from the electromagnetic to the hadronic channel. 
Muons are then produced in a second step in the hadronic sub-showers
that were initiated by a photonuclear interaction.
Changing the photonuclear cross-section influences the rate of
transferring energy to hadrons and, thus, the production of
secondary muons.

Also the position of depth of shower maximum is affected:
Increasing, for instance, the photonuclear cross-section
makes primary photon showers more similar to hadron-induced cascades.
Qualitatively, in this example, the $X_{\rm max}$ values would
become smaller for UHE photon showers,
i.e.~values closer to the average $X_{\rm max}$
for primary hadrons (see Figure~\ref{fig-xmax}) would emerge.

In the extreme case, an unconverted primary photon can undergo
photoproduction in the first interaction of the shower cascade.
This would result in a shower that hardly can be distinguished
from that of a hadron primary. The probability for such
a process is small, though. The electromagnetic Bethe-Heitler pair production
cross-section is about 500~mb whereas the photoproduction cross-section
is expected to be of the order 10~mb at $\simeq$10~EeV.
However, the LPM effect reduces the Bethe-Heitler cross-section
at ultra-high energy, and an uncertainty exists for extrapolating
the photonuclear cross-section.

%
\bfg[t]                     
\bc                         
\includegraphics[width=0.8\textwidth]{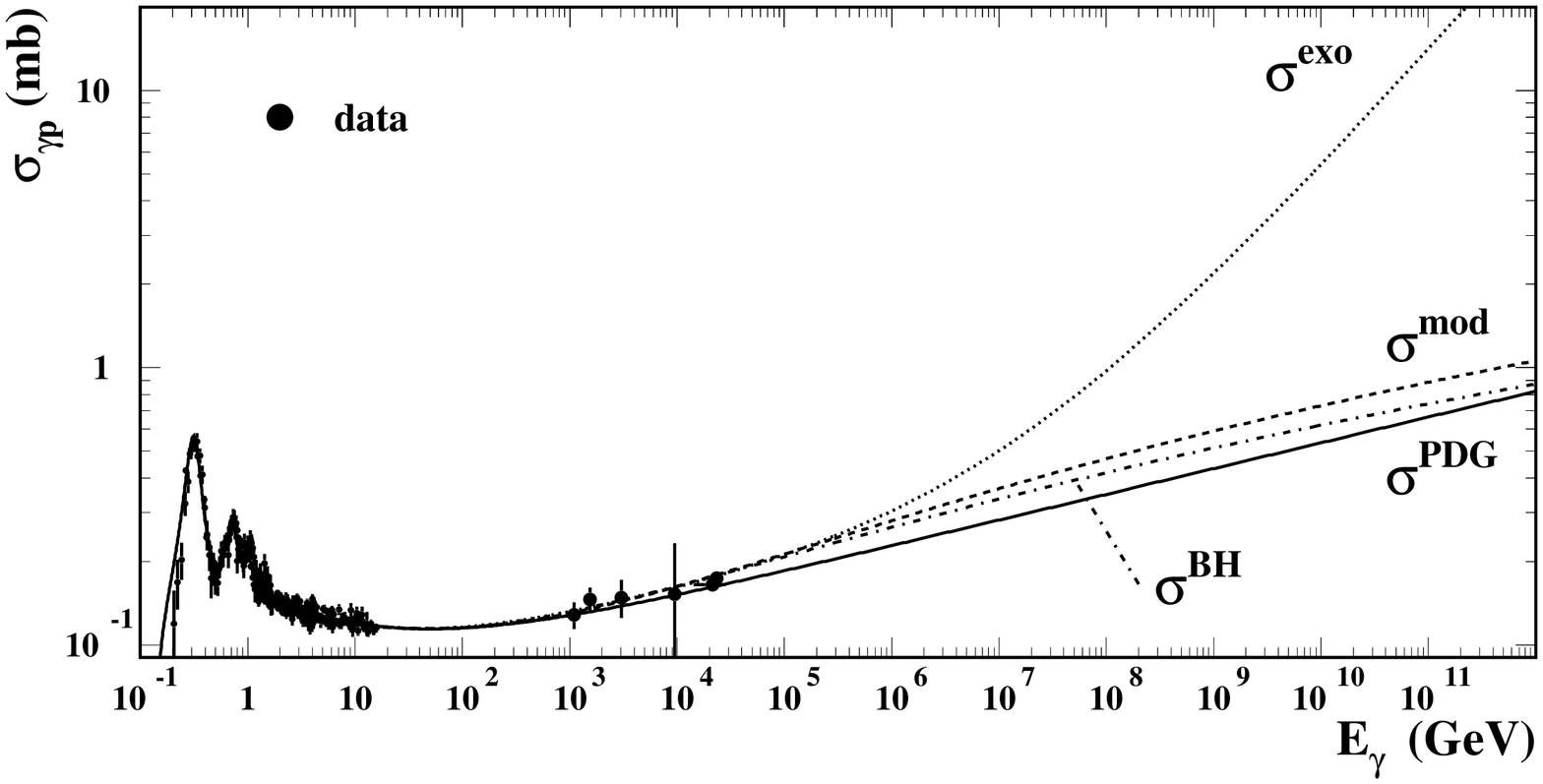}
\ec                         
\vspace{-2mm}
\caption{Data~\cite{pdg} and PDG extrapolation
($\sigma^{\rm PDG}$)~\cite{pdg,cud} of the
photonuclear cross-section $\sigma_{\gamma p}$.
Also shown are two parametrizations with larger
cross-sections at ultra-high energy,
denoted $\sigma^{\rm mod}$~\cite{bezbug} and
$\sigma^{\rm exo}$~\cite{donland} (see text),
and another recent fit $\sigma^{\rm BH}$~\cite{block}.
}
\label{fig-sigma}
\efg                        

In Figure~\ref{fig-sigma}, data and some published extrapolations for
the cross-section $\sigma_{\gamma p}$ are shown.
Large differences between the extrapolations are evident.
With respect to photon shower studies, several questions arise:
Is there a theoretical maximum for the allowed
increase of the cross-section (smaller values would make the
experimental limits to UHE photons even more severe)?
What is the impact on photon shower features when adopting different
cross-section extrapolations?
Is photon shower discrimination still possible even for very
rapidly rising cross-sections?
And, finally: If we were to observe photon showers, could this be
used in turn to constrain the photonuclear cross-section in an energy
range inaccessible at colliders?

To answer these questions, photon shower simulations were performed
for various cross-section assumptions with CORSIKA~\cite{corsika},
taking preshower formation~\cite{homola} and 
the LPM effect~\cite{lpm,corsika2} into account.
We consider here the extrapolation of the 
Particle Data Group, $\sigma^{\rm PDG}$~\cite{pdg,cud}, as a
baseline assumption. 
The most rapidly rising photoproduction cross-section is obtained
if the hard pomeron model of Donnachie and Landshoff~\cite{donland}
is extrapolated to high energy.
This model does not account for unitarity corrections that are expected
to become important at very high energy. Therefore, it should be
considered as a rather extreme extrapolation that could represent
an upper limit of possible low-energy data extrapolations.
In the following, we will refer to this cross-section extrapolation
as $\sigma^{\rm exo}$ (exotic).

{\it What is the impact on photon showers?}
The impact due to switching from $\sigma^{\rm PDG}$ to
$\sigma^{\rm exo}$ is quite
significant, increasing the number of ground muons by roughly
70$-$80\%~\cite{risse-prl}. The effect on $X_{\rm max}$ is a rather
moderate reduction
by $\simeq$30~g~cm$^{-2}$ for photons that formed a preshower before
entering the atmosphere~\cite{risse04}, see Figure~\ref{fig-eff}.
The reduction is found to be much larger (up to 100~g~cm$^{-2}$ or
more) for unconverted photons.

Recent theoretical work on a possible maximum cross-section
seems to indicate that cross-section values larger by a factor 2 or 
more than $\sigma^{\rm PDG}$ are very unlikely~\cite{strikman}.
A recent fit to the low-energy data, $\sigma^{\rm BH}$~\cite{block},
stays also close to $\sigma^{\rm PDG}$.
Using these cross-sections, the uncertainty of ultra-high energy
shower predictions is rather moderate.
For example, assuming the extrapolation 
$\sigma^{\rm mod}$~\cite{bezbug} (see Fig.~\ref{fig-sigma}) 
results in a change of $\simeq$7~g~cm$^{-2}$ in $X_{\rm max}$
and $\simeq$10\% in muon number.
Thus, it seems that a reasonable estimate of the present
uncertainty from extrapolating the photonuclear cross-section
is $\simeq$10~g~cm$^{-2}$ and $\simeq$15\% for $X_{\rm max}$ and
muon number, respectively.

%
\bfg[t]                     
\bc                         
\includegraphics[width=0.8\textwidth]{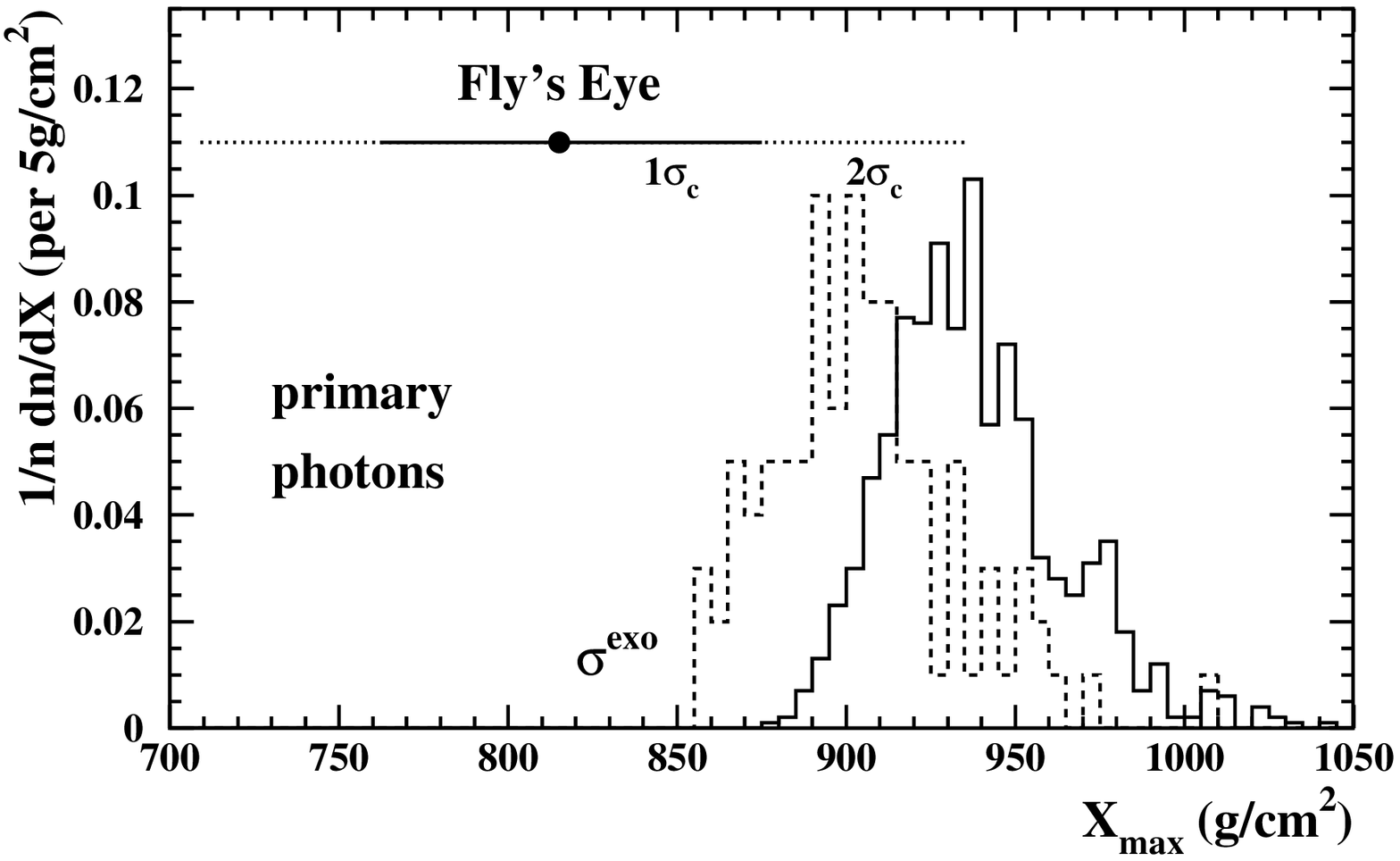}
\ec                         
\vspace{-2mm}
\caption{Shower maximum distribution of primary photons
compared to the reconstructed value of the Fly's Eye event.
The measured depth is shown with the $1\sigma$- and
$2\sigma$- uncertainty.
In addition to the simulations with standard photonuclear
cross-section,
results when assuming the extrapolation $\sigma^{exo}$
(see Fig.~\ref{fig-sigma}) are given~\cite{risse04}.}
\label{fig-eff}
\efg                        

{\it Is photon shower discrimination possible even for
rapidly rising cross-sections?}
Assuming an exotic scenario such as $\sigma^{\rm exo}$,
the discrimination power for photons is reduced, but still photons
are on average expected to have fewer muons (factor 2) and
larger $X_{\rm max}$ (50$-$100~g~cm$^{-2}$) compared to protons.
The separation to heavier nuclei is even larger.
Thus, if UHE photons exist, we have good chances to identify them.
In particular, the preshower effect can be used as a tool for photon
identification:
The detection of the characteristic dependence of $X_{\rm max}$
and muon number, or of observables related to these, on primary
direction and energy (both the average values of the shower observables
and their fluctuations)
would be an unambiguous signal of primary photons.

{\it Could the photonuclear cross-section be constrained at ultra-high
energy by air shower observations?}
When UHE photon showers are identified, their observed shower features
could in turn be used to constrain the photonuclear cross-section
at highest energies. The differences in $X_{\rm max}$ and muon
number when assuming $\sigma^{\rm PDG}$ or $\sigma^{\rm exo}$
can be determined even with relatively small event
statistics. Thus, it might be possible to convert the observed shower
features into an upper limit to the photonuclear cross-section.
It may also be useful to look for specific photon event classes:
For instance, photon events with a very delayed first interaction
(due to LPM suppression) would be much less frequent in the
$\sigma^{\rm exo}$ scenario, as they would start a hadronic cascade
at some point. The rate of deeply starting photon-like events
would constrain the photonuclear cross-section at these primary
energies.
\\

{\it Acknowledgments.}
It is a pleasure to thank the organizers, and in particular Jan Ridky,
for providing an inspiring conference atmosphere.
Helpful discussions with M.~Strikman, D.~Semikoz, M.M.~Block,
S.~Ostapchenko, and S.~Sarkar are kindly acknowledged.
This work was partially supported in Poland by the State Committee for
Scientific Research, grant No.~2P03B~11024, and in Germany by the DAAD,
grant No.~PPP~323.
One of the authors (MR) is supported by the
Alexander von Humboldt foundation.

%

\end{document}